\documentclass[conference,letterpaper]{IEEEtran}

\usepackage[utf8]{inputenc} 
\usepackage[T1]{fontenc}
\usepackage{url}
\usepackage{ifthen}
\usepackage{cite}
\usepackage[cmex10]{amsmath} 
\usepackage{amssymb, amsfonts}
\usepackage[pdftex]{graphicx}
\usepackage{cite}
\usepackage{cases}
\usepackage{algorithmic}
\usepackage{algorithm}

\DeclareMathOperator*{\Tr}{Tr}

\DeclareMathOperator*{\diag}{diag}
\DeclareMathOperator*{\Rank}{rank}

\newtheorem{theorem}{Theorem}

\newtheorem{define}{Definition}

\begin{document}
\title{Beamforming Design for Integrated Sensing and Communications Using Uplink-Downlink Duality} 

 \author{
 \IEEEauthorblockN{Kareem M. Attiah and Wei Yu} \IEEEauthorblockA{Electrical and Computer Engineering Department \\
                 University of Toronto, Canada\\
                    Emails: kattiah@ece.utoronto.ca, weiyu@ece.utoronto.ca}
 }

\maketitle


\begin{abstract}
This paper presents a novel optimization framework for
	beamforming design in integrated sensing and communication systems where a base station seeks to minimize the Bayesian Cramér-Rao bound of a sensing problem while satisfying quality of service constraints for the communication users.  
   Prior approaches formulate the design problem as a semidefinite program for which acquiring a beamforming solution is computationally expensive.  
   In this work, we show that the computational burden can be considerably alleviated. 
   To achieve this, we transform the design problem to a tractable form that not only provides a new understanding of Cramér-Rao bound optimization, but also allows for an uplink-downlink duality relation to be developed. Such a duality result gives rise to an efficient algorithm that enables the beamforming design problem to be solved at a much lower complexity as compared to the-state-of-the-art methods. 
\end{abstract}

\section{Introduction}
Next-generation wireless systems and their emerging applications are expected to offer high-throughput communication services as well as accurate sensing functionalities~\cite{liutcom202,ngu2022}. Integrated sensing and communication (ISAC) is a promising strategy which is anticipated to play a central role in meeting those demands. In contrast to traditional systems in which the sensing and communication operations are performed separately, ISAC systems offer better utilization of resources and network infrastructure. However, these benefits also come with the challenge of addressing more complicated signal processing tasks and hardware design.



This paper is motivated by the recent interest in beamforming design for the simultaneous operation of sensing and communications. Such problem is often posed as that of optimizing beamformers for a sensing-related metric, e.g., the Cramér-Rao bound (CRB) of an estimation problem, while guaranteeing certain quality-of-service, e.g., signal-to-interference-and-noise (SINR) constraints for the communication problem \cite{Liu2020joint, LiuF2022conf, LiuFCRB2022, Zhuinformation2023}. This optimization is challenging not only due to its nonconvexity but also because the objective function typically has rather complicated forms \cite{LiuFCRB2022, LiuF2022conf, Zhuinformation2023}. The best known strategy for tackling this problem relies on transforming the problem to the covariance domain, where the problem is convexified after dropping rank-one constraints using the semidefinite relaxation (SDR) technique. Despite the convexity of the transformed problem, solving the SDR problem is typically inefficient due to the need for lifting the solution space. Moreover, it often requires a randomization step whenever the SDR solution does not 
satisfy the rank-one constraint.

This paper provides a novel framework for solving this challenging
problem. The main idea is to transform the complicated ISAC problem into a simple form that involves minimizing the beamforming power subject to downlink SINR constraints. The main benefit of doing so is that the latter problem allows for an efficient solution based on uplink-downlink (UL-DL) duality \cite{yutransmitter2007, WieselLinear2006, Schubertsolution2004, rashidDL1998, rashidUL1998}. 


The task of transforming the ISAC problem into the desired form can be accomplished in two stages. First, we show that the optimization of a CRB objective can be viewed as that of maximizing the power in certain point of interest. This observation is related to a result in~\cite{Zhuinformation2023} for a specific angle estimation problem that we extend herein for an arbitrary vector parameter using a more general technique. The second step is to leverage Lagrangian theory in order to transform the resulting problem to that of minimizing the power subject to SINR constraints. Given the new formulation, we draw upon the existing theory in \cite{yutransmitter2007} to establish an UL-DL duality relation for the ISAC problem. Such duality result reveals that the same efficient algorithms developed for the classical communication problem~\cite{yutransmitter2007,rashidDL1998,rashidUL1998} are applicable here, with a few modifications. In contrast to the SDR approach, the proposed solution methodology is computationally efficient and does not lift the solution space. Finally, we present numerical results demonstrating the effectiveness of the proposed solution. 


\section{System Model}
\label{sec:sysmdl}
\subsection{ISAC Model and Performance Metrics}
\begin{figure}
    \centering    \includegraphics[width=0.342\textwidth]{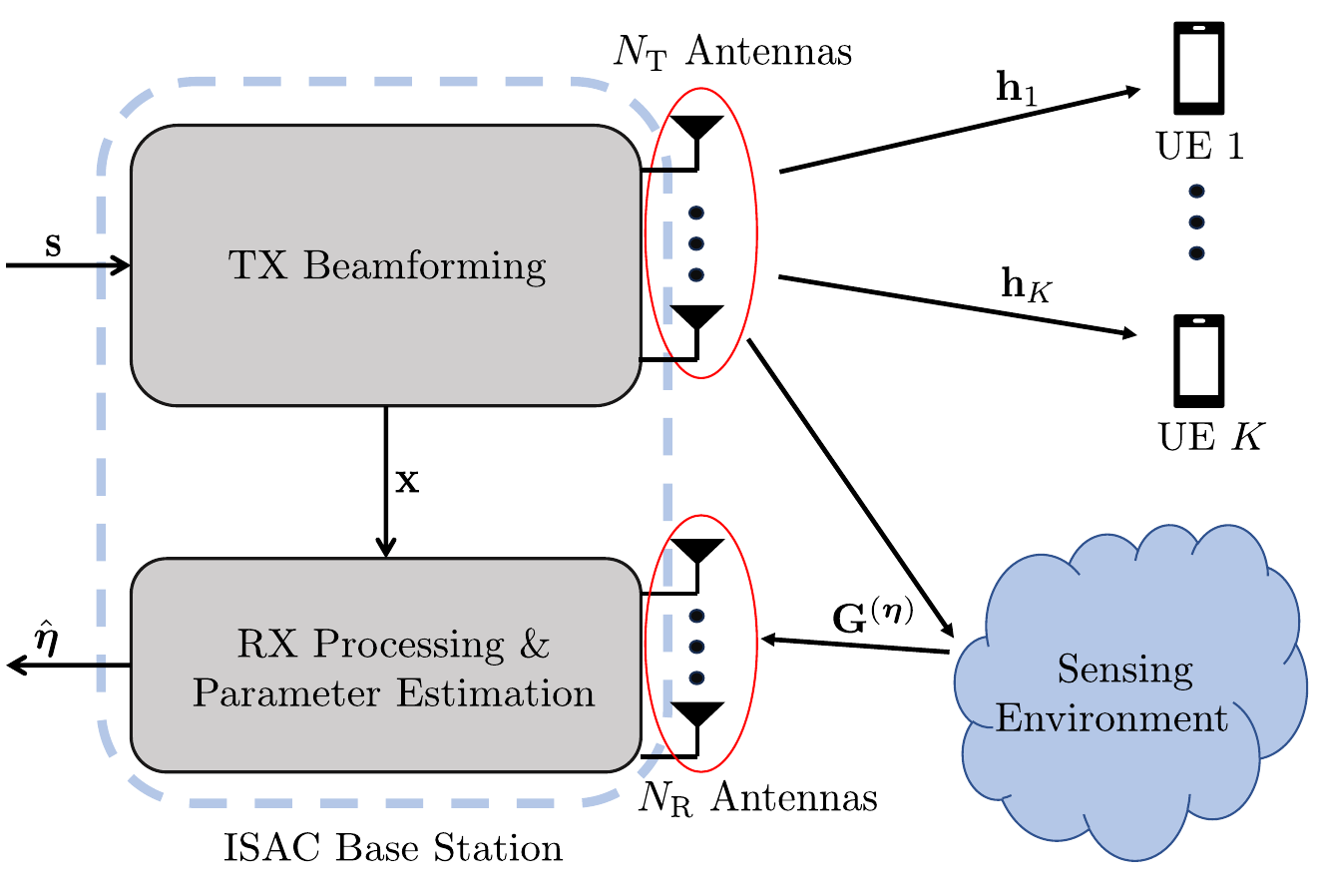}
     \caption{The ISAC system considered in this work. The BS seeks to serve $K$ communication users and learn some unknown vector parameter $\boldsymbol{\eta}$.}
     \label{fig:model}
\end{figure}

We consider the multi-user ISAC system illustrated in~\figurename~\ref{fig:model} over a coherence interval spanning $T$ symbols. 
In this setup, a base station (BS) equipped with co-located $N_\text{T}$ transmit antenna array and $N_\text{R}$ receive antenna array seeks to 
learn an unknown vector of real parameters $\boldsymbol{\eta} \in \mathbb{R}^{L}$. Simultaneously, it aims to
convey information to $K$ single-antenna communication users in the downlink. During the $t$-th symbol, the baseband transmitted signal $\mathbf{x}[t] \in \mathbb{C}^{N_\text{T}} $ follows a beamforming model given by
\begin{equation}
    \label{eq:BFMDL}
    \mathbf{x}[t] = \mathbf{V} \mathbf{s}[t] = \sum_k \mathbf{v}_k s_k[t], 
\end{equation}
where $\mathbf{V} \triangleq \left[\mathbf{v}_1, \ldots, \mathbf{v}_K\right] \in \mathbb{C}^{N_\text{T} \times K}$ is a matrix of beamforming vectors satisfying a total power constraint $\Tr\left(\mathbf{V} \mathbf{V}^\textsf{H} \right) \leq P$ and $\mathbf{s}[t]  = \left[s_1[t], \ldots, s_K[t]\right]^\textsf{T} \in \mathbb{C}^{K}$ is a zero-mean vector of communication symbols satisfying $\mathbb{E}\left\{ \mathbf{s}[t] \mathbf{s}^\textsf{H}[t]\right\} = \mathbf{I}_{K}$.  
Observe that the beamforming model~\eqref{eq:BFMDL} adopted herein is identical to that of conventional communication systems. Such model is particularly useful in scenarios where the length of the parameter to be estimated is short compared to the number of users (i.e., $L \ll K$), in which case $K$ degrees-of-freedom (DoFs) are enough to estimate the parameter~\cite{LiuFCRB2022, Liu2018MUMIMO}. In contrast, if $L$ is large, the beamforming model~\eqref{eq:BFMDL} should be augmented with an additional component dedicated to sensing~\cite{Liu2020joint,LiuF2022conf, HuaOptimal2023}. The extended model can exhibit better performance in this case due to the increased DoF, but at the cost of a more complicated design and increased RF-chain requirements. For simplicity, we restrict our attention to the model~\eqref{eq:BFMDL}.

In the $t$-th symbol, the BS transmits the waveform $\mathbf{x}[t]$ and subsequently observes the measurement vector $\mathbf{y}[t] \in \mathbb{C}^{N_\text{R}}$. In addition, each user observes a scalar baseband signal, denoted by $\tilde{y}_k[t]$ for the $k$-th user. Such received signals are given by
\begin{equation}
    \label{eq:radarmdl}
    \mathbf{y}[t] = 
    \mathbf{G}^{\left(\boldsymbol\eta \right)} 
    \mathbf{x}[t] + \mathbf{n}[t],
\end{equation}
and
\begin{equation}
        \label{eq:commmdl}
    \tilde{y}_k[t] =
    \mathbf{h}_k^\textsf{H}
    \mathbf{x}[t] + \tilde{n}_k[t], \quad \forall k \in \left\{1, \ldots, K\right\},
\end{equation}
where $\mathbf{h}_k \in \mathbb{C}^{N_\text{T}}$ is the $k$-th user channel and $\mathbf{n}[t] \sim \mathcal{CN}(0, \sigma^2 \mathbf{I}_{N_\text{R}})$ and $\tilde{n}_k[t] \sim \mathcal{CN}\left(0, \sigma^2\right)$ are noise terms whose variances are normalized to be the same without loss of generality.
The matrix $\mathbf{G}^{(\boldsymbol{\eta})} \in \mathbb{C}^{N_\text{R} \times N_\text{T}}$ models the ``round-trip" channel between the transmit and receive antenna arrays. Observe that $\boldsymbol{\eta}$ is the set of underlying parameters of the sensing channel. This model encompasses a wide variety of practical scenarios (e.g., angle-of-arrival estimation of a radar target). Finally, we assume that the communication channel $\mathbf{H} \triangleq \left[\mathbf{h}_1, \ldots, \mathbf{h}_K\right] \in \mathbb{C}^{N_\text{T} \times K}$ is perfectly known at the BS. 

We aim to design the beamforming matrix $\mathbf{V}$ so that certain performance goals are met for communication and sensing. For sensing, we assume a prior distribution on $\boldsymbol{\eta}$ and adopt the Bayesian CRB (BCRB) as a measure of performance. Unlike the classical CRB used in previous works~\cite{LiR2008ange,LiuF2022conf, LiuFCRB2022,Huamimo2023,bekkerTarget2006} which depends on the actual value of the unknown parameter $\boldsymbol{\eta}$, the BCRB provides a bound on the mean-squared error (MSE) averaged over the prior. 
For an estimator $\hat{\boldsymbol{\eta}}$ 
we have $\mathbb{E}\{ \left( \boldsymbol{\eta} - \hat{\boldsymbol{\eta}}\right) \left( \boldsymbol{\eta} - \hat{\boldsymbol{\eta}} \right)^\textsf{T} \} \succcurlyeq \mathbf{J}^{-1}_{\mathbf{V}}$, where $\succcurlyeq$ denotes inequality with respect to the positive semidefinite (PSD) cone and $\mathbf{J}_{ \mathbf{V}} \in  \mathbb{R}^{L \times L}$ is the Bayesian Fisher information matrix (BFIM). For the additive Gaussian noise channel model~\eqref{eq:radarmdl}, the elements of BFIM can be expressed as follows \cite{kay1993fundamentals, Huleihel2013optimal}:
\begin{equation} \label{eq:FIM_elements}
\left[\mathbf{J}_{\mathbf{V}}\right]_{ij} = \frac{2T}{\sigma^2} \Re\left(\Tr\left( \mathbb{E} \left\{\dot{\mathbf{G}}_i^\textsf{H} \dot{\mathbf{G}}_j\right\} \mathbf{V} \mathbf{V}^\textsf{H} \right) \right) + [\mathbf{C}]_{ij}
\end{equation}
where $\dot{\mathbf{G}}_i \triangleq \tfrac{\partial \mathbf{G}^{\left(\boldsymbol\eta \right)} }{\partial \eta_i}$ and $\mathbf{C} \in \mathbb{R}^{L \times L}$ is a matrix whose elements are given by $[\mathbf{C}]_{ij} \triangleq - \mathbb{E} \{\tfrac{\partial^2 \log f(\boldsymbol{\eta}) }{\partial \eta_i \partial \eta_j}\}$. Observe that $\mathbf{C}$ 
does not depend on $\mathbf{V}$. 
Our aim is to minimize the trace of the inverse of the BFIM
\begin{equation} \label{eq:sensing_metric}
    \mu(\mathbf{V}) \triangleq \Tr \left( \mathbf{J}^{-1}_{ \mathbf{V}}\right).
\end{equation}
The above metric constitutes a lower bound on the sum of average MSEs across the individual elements of $\boldsymbol{\eta}$. The main advantage of the metric~\eqref{eq:sensing_metric} is that it abstracts the receiver design since the bound holds for any estimator.

For communications, most existing works on ISAC assume a performance metric of satisfying certain rate constraints for the users~\cite{Liu2020joint,LiuF2022conf, LiuFCRB2022}. This paper also follows the same formulation. Since the rate is a monotone function of the signal-to-interference-and-noise-ratio (SINR), the rate requirements can be described as constraints on the SINR, i.e., 
\begin{equation*}
    \label{eq:SINR}
    \text{SINR}^{\text{DL}}_k(\mathbf{V}) \triangleq 
\frac{\left|\mathbf{h}_k^\textsf{H} \mathbf{v}_{k} \right|^2}{\sum_{i \neq k} \left| \mathbf{h}_k^\textsf{H} \mathbf{v}_{i}\right|^2 + \sigma^2} \geq \gamma_k, \quad \forall k, 
\end{equation*}
where $\gamma_k$ denotes the SINR threshold for the $k$-th user.


\subsection{Problem Formulation and Solution Strategy}

Based on the previous considerations, the beamforming design problem can be mathematically stated as follows: 
\begin{subequations}\label{prob:generalCase}
    \begin{align}
    \underset{\mathbf{V}}{\mathrm{minimize}} ~~~~ &  \Tr \left( \mathbf{J}^{-1}_{\mathbf{V}}\right) 
    \label{eq:general_obj} \\
            \mathrm{subject \ to}  ~~~ & \frac{\left|\mathbf{h}_k^\textsf{H} \mathbf{v}_k \right|^2}{\sum_{i \neq k} \left|\mathbf{h}_k^\textsf{H} \mathbf{v}_i \right|^2 + \sigma^2} \geq \gamma_k, \quad \forall k, \label{eq:SINR_const} \\
            &\Tr \left(\mathbf{V} \mathbf{V}^\textsf{H} \right) \leq P. \label{eq:pw_const}
    \end{align}
\end{subequations}
We make the assumption that the constraints of the above problem are feasible. 
Problem~\eqref{prob:generalCase} is challenging to solve for a number of reasons. 
First, the objective is a complicated function to compute let alone optimize. Second, neither the objective nor the feasible set is convex in $\mathbf{V}$. 

One possible strategy for solving \eqref{prob:generalCase} is SDR. For example,
the work~\cite{LiuFCRB2022} considers a special case of \eqref{prob:generalCase} in which the sensing task is that of estimating the angle-of-arrival for an unknown target. For this simpler case, the authors of~\cite{LiuFCRB2022} show that an SDR strategy 
can effectively relax the problem and transform it into a convex form. 
Despite the convexity,
the SDR approach suffers from a number of drawbacks. First, the SDR solution is not guaranteed to satisfy the rank constraint. If the SDR solution is not rank one, we must employ an additional, and often suboptimal, algorithm 
to recover a rank-one solution. Secondly, it lifts the optimization space from ${N}_\text{T}K$ complex dimensions to $N_\text{T}^2K$ complex dimensions. Consequently, the interior-point method requires 
$\mathcal{O}(N_\text{T}^{6} K^3)$ per iteration to compute a search direction~\cite{boyd2004convex}. This order of complexity is infeasible in many practical scenarios. 

A key contribution of this paper is to develop an efficient optimization method that does not require lifting the problem dimensionality. 
We accomplish this by establishing a connection between \eqref{prob:generalCase} and the following classical problem
\begin{subequations}\label{prob:dl_comm}
    \begin{align}
            \underset{\mathbf{V}}{\mathrm{minimize}} ~~~~~& \sum_k \| {\mathbf{v}}_k \|^2 \label{eq:comm_obj} \\
            \mathrm{subject \ to}  ~~~~ & \frac{\left|\mathbf{h}_k^\textsf{H} \mathbf{v}_k \right|^2}{\sum_{i \neq k} \left|\mathbf{h}_k^\textsf{H} \mathbf{v}_i \right|^2 + \sigma_\text{c}^2} \geq \gamma_k, \quad \forall k.
    \end{align}
\end{subequations}
Problem~\eqref{prob:dl_comm} arises in the context of downlink MIMO beamforming where the goal is to ensure reliable communication for the users (by constraining the SINRs) while minimizing the transmit power at the BS. 
In~\cite{rashidDL1998}, a duality notion is observed that allows the downlink problem~\eqref{prob:dl_comm} to be transformed to a virtual uplink problem, a result now commonly referred to as UL-DL duality~\cite{rashidDL1998, Schubertsolution2004, WieselLinear2006, yutransmitter2007}. The key advantage of such transformation is that the uplink problem is more computationally efficient to solve than its downlink counterpart. 


\section{Problem Reformulation as Downlink Power Maximization Problem}
We begin by showing that the ISAC problem~\eqref{prob:generalCase} can be expressed as a simpler problem that involves  maximizing a quadratic term subject to the constraints~\eqref{eq:SINR_const} and~\eqref{eq:pw_const}. In particular, we aim to rewrite problem~\eqref{prob:generalCase} as a sequence of maximization problems in the following form
\begin{subequations} \label{prob:qcqp}
\begin{align} 
	\underset{\mathbf{V}}{
		\mathrm{maximize}} \quad &  \Tr{\left( \mathbf{Q}  \mathbf{V} \mathbf{V}^\textsf{H} \right)} \label{eq:qcqp_obj} \\ 
	\mathrm{subject \ to} \quad  & \frac{\left|\mathbf{h}_k^\textsf{H} \mathbf{v}_k \right|^2}{\sum_{i \neq k} \left|\mathbf{h}_k^\textsf{H} \mathbf{v}_i \right|^2 + \sigma_\text{c}^2} \geq \gamma_k, \quad \forall k. \label{eq:qcqp_constraints} \\
	     & \Tr \left(\mathbf{V} \mathbf{V}^\textsf{H} \right) \leq P.  \label{eq:qcqp_power_constraint}
\end{align}
\end{subequations}
 where 
 $\mathbf{Q} \in \mathbb{C}^{N_\text{T} \times N_\text{T}}$ is a PSD matrix. Problem~\eqref{prob:qcqp} has the property that it has zero duality gap. This follows from the observation that a rank-one solution must exist for problems of this particular type; see~\cite{huangrank2010}.  
 We now proceed to transform problem~\eqref{prob:generalCase} into this desired form.


\begin{theorem}
\label{thm:minimax_SpecialCase} 
     If problem~\eqref{prob:generalCase} is feasible, it is equivalent to 
    \begin{align} \label{prob:min_max_specialcase_1}
	    \underset{\boldsymbol{\beta} \in \mathbb{R}^{L \times L}}{\mathrm{maximize}}~~\min_{\mathbf{V} \in \mathcal{V} } ~~ \sum_{\ell = 1}^L \left( 2 \boldsymbol{\boldsymbol{\beta}}_\ell^\textsf{T} \mathbf{e}_\ell - \boldsymbol{\beta}_\ell^\textsf{T} \mathbf{J}_{\mathbf{V}} \boldsymbol{\beta}_\ell \right) 
\end{align}
where $\boldsymbol{\beta} \triangleq [\boldsymbol{\boldsymbol{\beta}}_1, \ldots, \boldsymbol{\boldsymbol{\beta}}_L] \in \mathbb{R}^{L \times L}$,   
$\mathbf{e}_\ell$ is the $\ell$-th column of $\mathbf{I}$, 
and $\mathcal{V}$ denotes 
	the constraints \eqref{eq:qcqp_constraints}-\eqref{eq:qcqp_power_constraint}. Further, problem~\eqref{prob:min_max_specialcase_1} can be re-written as 
   \begin{align} \label{prob:min_max_specialcase}
	   \underset{\boldsymbol{\beta} \in \mathbb{R}^{L \times L}}{\mathrm{maximize}}~\min_{\mathbf{V} \in \mathcal{V}} \left[ \left( \sum_{\ell = 1}^L 2  \boldsymbol{\boldsymbol{\beta}}_\ell^\textsf{T} \mathbf{e}_\ell -\boldsymbol{\beta}_{\ell}^\textsf{T} \mathbf{C} \boldsymbol{\beta}_{\ell} \right) - \Tr{\left(\mathbf{Q}_{\boldsymbol{\beta}} \mathbf{V}\mathbf{V}^H \right)} \right]
\end{align}
where $
\mathbf{Q}_{\boldsymbol{\beta}} \triangleq \frac{2T}{\sigma^2}  \sum_{\ell = 1}^L  \mathbb{E}\{\tilde{\mathbf{G}}_{\boldsymbol{\beta_\ell}}^\textsf{H}\tilde{\mathbf{G}}_{\boldsymbol{\beta_\ell}}\},
$ and $\tilde{\mathbf{G}}_{\boldsymbol{\beta_\ell}} \triangleq \sum_i [\boldsymbol{\boldsymbol{\beta}}_\ell]_i \dot{\mathbf{G}}_i$.
In particular, there exists $\boldsymbol{\boldsymbol{\beta}}^*$ such that by setting $\mathbf{Q} = \mathbf{Q}_{\boldsymbol{\beta}^*}$ in \eqref{prob:qcqp}, we can obtain the optimal beamformers in problem \eqref{prob:generalCase}.
\begin{IEEEproof} 
The first part of Theorem~\ref{thm:minimax_SpecialCase} 
	is related to a similar result in \cite{Zhuinformation2023} but for a specific angle estimation problem and using an extra sensing beamformer, so its formulation is different. Further, the proof technique in~\cite{Zhuinformation2023} relies on a quadratic transform for fractional programming~\cite{shenFP2018},
while we provide an alternative proof based on Schur complement. 
For notational simplicity, we suppress the dependence on $\mathbf{V}$ in $\mathbf{J}$. We start with the convex relaxation, i.e., SDR of problem~\eqref{prob:generalCase}%
\begin{subequations}\label{prob:general_SDP_main_text}
    \begin{align}
    \underset{\mathbf{R}_1,\ldots, \mathbf{R}_K}{\mathrm{minimize}} ~~&  \Tr \left( \mathbf{J}^{-1} \right) 
    \label{eq:sdp_obj} \\
    \mathrm{subject \ to}   ~& \frac{1}{\gamma_k}  \mathbf{h}_k^\textsf{H} \mathbf{R}_k \mathbf{h}_k - \sum_{i \neq k}  \mathbf{h}_k^\textsf{H} \mathbf{R}_i \mathbf{h}_k \geq \sigma^2 \label{eq:sdp_const_1} \\
& \sum_k \Tr \left( \mathbf{R}_k \right) \leq P, \quad \mathbf{R}_k \succcurlyeq 0 ~~\forall k, \label{eq:sdp_const_2}
    \end{align}
\end{subequations}
where $\mathbf{R}_k \triangleq \mathbf{v}_k \mathbf{v}_k^\textsf{H}$. 
Note that by $\mathbf{J}$ is an affine function of $\mathbf{R}_k$ by \eqref{eq:FIM_elements}, so the above is a convex optimization problem.

We use the following trick to express the objective~\eqref{eq:sdp_obj} using Schur complement \cite{vandenberghe1998determinant}\cite{Huleihel2013optimal, kokke2023} as follows:
\begin{subequations}\label{prob:trace_inv}
\begin{align}
     \Tr \left( \mathbf{J}^{-1} \right) = \min_{d_1, \ldots, d_L} ~~ &  \sum_\ell d_\ell \\
     \mathrm{s.t.\ }  ~~~~ & \left[ \begin{matrix} \mathbf{J} & e_\ell \\ e_\ell^\textsf{T} & d_\ell \end{matrix} \right] \succcurlyeq 0, ~\forall \ell. \label{eq:SDP_const_main_text}
\end{align}
\end{subequations}
Combining~\eqref{prob:general_SDP_main_text} with~\eqref{prob:trace_inv}, we obtain
\begin{subequations}\label{prob:general_SDP2_main_text}
    \begin{align}
    \underset{ \mathbf{R}_1,\ldots, \mathbf{R}_K, d_1, \ldots, d_L}{\mathrm{minimize}} ~~ &  \sum_\ell d_\ell  \\
    \mathrm{subject \ to}  ~~~~ & \eqref{eq:sdp_const_1}, \eqref{eq:sdp_const_2}, \eqref{eq:SDP_const_main_text}
    \end{align}
\end{subequations} 
Problem~\eqref{prob:general_SDP2_main_text} is convex with strong duality. 
We formulate its dual problem with respect to \eqref{eq:SDP_const_main_text}. 
Let $\tilde{\mathbf{B}}_1, \ldots, \tilde{\mathbf{B}}_L \in \mathbb{R}^{(L + 1) \times (L + 1)}$ denote the dual variables associated with~\eqref{eq:SDP_const_main_text} 
\begin{equation}
\tilde{\mathbf{B}}_\ell \triangleq \left[\begin{matrix}
        \mathbf{B}_\ell & -\boldsymbol{\beta}_\ell \\
        -\boldsymbol{\beta}_\ell^\textsf{T} & b_\ell 
    \end{matrix} \right]   \succcurlyeq 0,
\end{equation}
with $\mathbf{B}_\ell \in \mathbb{R}^{L \times L}$, $ \boldsymbol{\beta}_\ell \in \mathbb{R}^{L}$, and $b_\ell \in \mathbb{R}$. The dual problem is
\begin{align*}
\underset{\tilde{\mathbf{B}}_1, \ldots, \tilde{\mathbf{B}}_L}{\mathrm{maximize}}\min_{\underset{d_1, \ldots, d_L}{(\mathbf{R}_1,\ldots, \mathbf{R}_K}) \in \mathcal{R}}   \sum_\ell d_\ell (1 - b_\ell) + 2 e_\ell^\textsf{T} \boldsymbol{\beta}_\ell 
     -\Tr(\mathbf{B}_\ell\mathbf{J}).
    \end{align*}
where $\mathcal{R}$ denotes the constraints~\eqref{eq:sdp_const_1}-\eqref{eq:sdp_const_2}.  
By minimizing the above over $d_\ell$, we conclude 
$b^*_\ell = 1$ for all $\ell$. Then, the dual problem becomes
\begin{align}\label{eq:max_min_R}
\underset{\mathbf{B}_\ell \succcurlyeq \boldsymbol{\beta}_\ell \boldsymbol{\beta}_\ell^\textsf{T}, \forall \ell}{\mathrm{maximize}}~~\min_{(\mathbf{R}_1,\ldots, \mathbf{R}_K) \in \mathcal{R}}    \sum_\ell  2 e_\ell^\textsf{T} \boldsymbol{\beta}_\ell 
     -\Tr(\mathbf{B}_\ell\mathbf{J}),
    \end{align}
    where the condition $\mathbf{B}_\ell \succcurlyeq \boldsymbol{\beta}_\ell \boldsymbol{\beta}_\ell^\textsf{T}$ follows
    because $\tilde{\mathbf{B}}_\ell \succcurlyeq 0$ and $b_\ell = 1$, so the Schur complement must satisfy $\mathbf{B}_\ell - \boldsymbol{\beta}_\ell \boldsymbol{\beta}_\ell^\textsf{T} \succcurlyeq 0$. Now, since strong duality holds, the primal optimal $\mathbf{R}^*_1, \ldots, \mathbf{R}^*_K$ and dual optimal $\mathbf{B}_1^*, \ldots, \mathbf{B}^*_L, \boldsymbol{\beta}^*_1, \ldots, \boldsymbol{\beta}^*_L$ is a saddle-point solution of the max-min problem. By interchanging $\min$ and $\max$, we can optimize over $\mathbf{B}_\ell$. In particular, for any fixed $\mathbf{R}_1, \ldots, \mathbf{R}_K$, we have
    \begin{equation}
        \mathbf{B}_\ell \succcurlyeq \boldsymbol{\beta}_\ell \boldsymbol{\beta}_\ell^\textsf{T} \Rightarrow \Tr(\mathbf{B}_\ell \mathbf{J})   \geq \Tr(\boldsymbol{\beta}_\ell \boldsymbol{\beta}_\ell^\textsf{T} \mathbf{J}).
    \end{equation}
	which holds because $\mathbf{J}$ is positive definite.  Thus, the optimal $\mathbf{B}_\ell$ must be $\mathbf{B}_\ell = \boldsymbol{\beta}_\ell \boldsymbol{\beta}^\textsf{T}_\ell$ at the saddle-point. In other words, the saddle-point in \eqref{eq:max_min_R} can be alternatively written as
\begin{align}\label{prob:general_SDP4_main_text}
	\underset{\boldsymbol{\beta} \in \mathbb{R}^{L \times L}}{\mathrm{maximize}}~~\min_{{(\mathbf{R}_1,\ldots, \mathbf{R}_K) \in \mathcal{R}}}   \sum_\ell \left( 2 e_\ell^\textsf{T} \boldsymbol{\beta}_\ell 
	-\ \Tr(\boldsymbol{\beta}_\ell \boldsymbol{\beta}_\ell^\textsf{T} \mathbf{J}) \right).
\end{align}
We now show that there exists a rank-one solution in which $\Rank (\mathbf{R}_1^*) = \cdots = \Rank (\mathbf{R}_K^*) = 1$. To see this, note that based on \eqref{eq:FIM_elements}, we can rewrite
	\begin{equation}
		 \sum_\ell \Tr(\boldsymbol{\beta}_\ell \boldsymbol{\beta}_\ell^\textsf{T} \mathbf{J}) =
		\sum_\ell \boldsymbol{\beta}_\ell^\textsf{T} \mathbf{J} \boldsymbol{\beta}_\ell = \sum_\ell \boldsymbol{\beta}_\ell^\textsf{T} \mathbf{C} \boldsymbol{\beta}_\ell + \sum_k \Tr(\mathbf{Q}_{\boldsymbol{\beta}} \mathbf{R}_k), 
	\end{equation}
	where $\mathbf{Q}_{\boldsymbol{\beta}}$ is as defined in the theorem statement.
In other words, for any fixed $\boldsymbol{\beta}$, the inner minimization in~\eqref{prob:general_SDP4_main_text} has the form of the SDR of a problem~\eqref{prob:qcqp}. But the SDR of \eqref{prob:qcqp} is tight by results of \cite{huangrank2010}. 
This means that there exists a set of rank-one solution $\mathbf{R}_1^*, \ldots \mathbf{R}_K^*$,
i.e., $\mathbf{R}_k^* = \mathbf{v}_k^* {\mathbf{v}_k^*}^\textsf{H}$,
for the inner minimization problem in~\eqref{prob:general_SDP4_main_text},
for any $\boldsymbol{\beta}$ and in particular for $\boldsymbol{\beta} = \boldsymbol{\beta}^*$. 
This implies that
(\ref{prob:general_SDP_main_text}), the SDR of (\ref{prob:generalCase}), is equivalent to \eqref{prob:generalCase} itself, and 
$\mathbf{V}^* \triangleq [\mathbf{v}_1^*, \ldots, \mathbf{v}_K^*]$ is the solution of the original ISAC problem~\eqref{prob:generalCase}. This shows that problem~\eqref{prob:generalCase} is equivalent to 
problems \eqref{prob:min_max_specialcase_1} and \eqref{prob:min_max_specialcase}.
\end{IEEEproof}
\end{theorem} 

Theorem~\ref{thm:minimax_SpecialCase} allows us to express the ISAC problem~\eqref{prob:generalCase} in a more tractable form. In addition, it offers the following interesting interpretation. Consider the max-min problem~\eqref{prob:min_max_specialcase}. By construction, the matrix $\mathbf{Q}_{\boldsymbol{\beta}}$ is PSD, so the trace term in \eqref{prob:min_max_specialcase} can be interpreted as a sum of beamforming powers measured with respect to the matrix $\mathbf{Q}_{\boldsymbol{\beta}}$. Thus, for a fixed $\boldsymbol{\boldsymbol{\beta}}$, the inner problem in~\eqref{prob:min_max_specialcase} can be viewed as maximizing the beamforming power at certain point of interest. Changing $\boldsymbol{\boldsymbol{\beta}}$ corresponds to changing the point of interest. Theorem~\ref{thm:minimax_SpecialCase} essentially states that there exists a judicious choice $\boldsymbol{\boldsymbol{\beta}}^*$ for which the problem of minimizing the CRB is equivalent to that of maximizing the power along a certain direction.

\section{Uplink-Downlink Duality for ISAC}
 We now show that the ISAC problem admits an alternative formulation in terms of an uplink problem. 
 The reason that an UL-DL duality relation exists for ISAC beamforming 
is that problems~\eqref{prob:dl_comm} and~\eqref{prob:qcqp} are closely related. 
On one hand, problem~\eqref{prob:qcqp} (or the ISAC version when $\mathbf{Q} = \mathbf{Q}_{\boldsymbol{\beta}^*}$) seeks to maximize the transmit power in certain directions under a total power constraint. On the other hand, problem~\eqref{prob:dl_comm} aims to minimize the transmit power. 
Consider the inner problem in~\eqref{prob:min_max_specialcase} for a given $\boldsymbol{\beta}$. We introduce the Lagrangian function 
\begin{equation*}
    \mathcal{L}_{\lambda, \boldsymbol{\beta}}\left( \mathbf{V}\right) = \sum_k \mathbf{v}_k^\textsf{H} \left(\lambda \mathbf{I} - \mathbf{Q}_{\boldsymbol{\beta}}  \right)  \mathbf{v}_k - \lambda P + \sum_{\ell = 1}^L 2  \boldsymbol{\boldsymbol{\beta}}_\ell^\textsf{T} \mathbf{e}_\ell -\boldsymbol{\beta}_{\ell}^\textsf{T} \mathbf{C} \boldsymbol{\beta}_{\ell}
\end{equation*}
which gives rise to the following dual problem
\begin{subequations}\label{prob:partial_dual}
\begin{align} 
\underset{\boldsymbol{\beta}, \lambda \geq 0 }{\mathrm{maximize}}~~~\min_{\mathbf{V}}~~~~~&\mathcal{L}_{\lambda, \boldsymbol{\beta}}\left( \mathbf{V}\right) \\ 
 \mathrm{s.t.\ }  ~~~~~ & \text{SINR}^{\text{DL}}_k(\mathbf{V}) \geq \gamma_k, ~~ \forall k.
\end{align}
\end{subequations}
Because strong duality holds for the inner problem in~\eqref{prob:min_max_specialcase}, an optimal beamforming solution of problem~\eqref{prob:min_max_specialcase} (or problem~\eqref{prob:generalCase}) must also be a minimizer of $\mathcal{L}_{\lambda^*, \boldsymbol{\beta}^*}\left( \cdot\right)$. 

Now, consider the inner problem in~\eqref{prob:partial_dual} for fixed $(\lambda, \boldsymbol{\beta})$: 
\begin{subequations}\label{prob:fixed_lambda}
\begin{align}
\underset{\mathbf{V}}{\mathrm{minimize}}~~~~&  \sum_k \mathbf{v}_k^\textsf{H} \left(\lambda \mathbf{I} - \mathbf{Q}_{\boldsymbol{\beta}}  \right)  \mathbf{v}_k  \label{prob:fixed_lambda_obj} \\ 
 \mathrm{subject \; to}  ~~~& \text{SINR}^{\text{DL}}_k(\mathbf{V}) \geq \gamma_k, ~~ \forall k. \label{prob:fixed_lambda_dl_sinr}
\end{align}
\end{subequations}
For suitable choice of $\lambda$ and $\boldsymbol{\beta}$, this problem can be viewed as a downlink communication problem, similar to the classical problem~\eqref{prob:dl_comm}. For example, if 
$\lambda \mathbf{I} -\mathbf{Q}_{\boldsymbol{\beta}} \succcurlyeq 0$, then \eqref{prob:fixed_lambda_obj} represents a sum of beamforming powers measured with respect to $\lambda \mathbf{I} -\mathbf{Q}_{\boldsymbol{\beta}}$. 
A natural question is the following. For what values of $(\lambda, \boldsymbol{\beta})$ does problem~\eqref{prob:fixed_lambda} admit a similar interpretation as problem~\eqref{prob:dl_comm}? This motivates the following definition.
\begin{define}\label{define:admissibility}
The pair $\lambda, \boldsymbol{\beta}$ is said to be admissible if %
    $\sum_k \mathbf{v}_k^\textsf{H} \left( \lambda \mathbf{I} - \mathbf{Q}_{\boldsymbol{\beta}} \right)\mathbf{v}_k  \geq 0$
for all $\mathbf{V}$ satisfying~\eqref{prob:fixed_lambda_dl_sinr}. \hfill \IEEEQED
\end{define}

Admissibility of $\lambda, \boldsymbol{\beta}$ implies that the matrix $\lambda \mathbf{I} - \mathbf{Q}_{\boldsymbol{\beta}}$ behaves as a ``weak" PSD matrix. It may not be PSD globally but behaves like one over the feasible region of SINRs. 
This also implies that the objective function~\eqref{prob:fixed_lambda_obj} is a convex function when restricted to $\mathbf{v}_k$'s that satisfy the set of SINRs, but can be nonconvex outside such set. 

Now, let $\mathcal{A}$ be the set of all admissible pairs. The next result establishes an UL-DL relation for problem~\eqref{prob:fixed_lambda} if $(\lambda, \boldsymbol{\beta}) \in \mathcal{A}$. 
\begin{theorem}
\label{thm:ULDL_duality}
Problem~\eqref{prob:fixed_lambda} is bounded below if and only if $(\lambda, \boldsymbol{\beta}) \in \mathcal{A}$. In addition, whenever $(\lambda, \boldsymbol{\beta}) \in \mathcal{A}$, the matrix
\begin{equation}\label{eq:opt_V_fixed}
        \mathbf{V}^*_{\lambda, \boldsymbol{\beta}} \triangleq \mathbf{U}^*_{\lambda, \boldsymbol{\beta}} \mathbf{p}_{\lambda, \boldsymbol{\beta}}^*, 
\end{equation}
        is an optimal solution of the problem. Here, $\mathbf{p}^*_{\lambda, \boldsymbol{\beta}} \triangleq \sigma^2 ( \mathbf{D}_{\mathbf{U}_{\lambda, \boldsymbol{\beta}}^*} - \mathbf{F}_{\mathbf{U}_{\lambda, \boldsymbol{\beta}}^*})^{-1} \mathbf{1} \in \mathbb{R}^{K}$ is a vector of downlink powers
     and $\mathbf{U}^*_{\lambda, \boldsymbol{\beta}} \in \mathbb{C}^{N_\text{T} \times K}$ is a matrix of normalized beamforming directions obtained by solving the uplink problem
\begin{subequations} \label{prob:UL_problem_fixed_lambda}
\begin{align} 
\underset{\mathbf{U}, \; \mathbf{z}}{\mathrm{minimize}}  ~~& \sigma^2  \mathbf{1}^\textsf{T} \mathbf{z}   \\
    \mathrm{subject \ to} ~& \frac{z_k \left|\mathbf{h}_k^\textsf{H} \mathbf{u} \right|^2}{\sum_{i \neq k}  z_i \left| \mathbf{h}_i^\textsf{H}\mathbf{u}\right|^2 +  \mathbf{u}^\textsf{H} 
    \left( \lambda \mathbf{I} - \mathbf{Q}_{\boldsymbol{\beta}} \right)
    \mathbf{u}}  \geq \gamma_k,   \label{eq:ul_sinr} \\
    &\rho_\text{max} \left( \mathbf{D}_\mathbf{U}^{-1} \mathbf{F}_\mathbf{U} \right) < 1, \label{eq:spectral_rad}
 \\
      &  \| \mathbf{u}_k \|_2 = 1, \quad \forall k. \label{eq:normality}
\end{align}
\end{subequations}
    In problem~\eqref{prob:UL_problem_fixed_lambda}, $\mathbf{U} \triangleq \left[\mathbf{u}_1, \ldots, \mathbf{u}_K\right]$ represents a matrix of combining vectors and $\mathbf{z} \triangleq \left[z_1, \ldots, z_K\right]^\textsf{T} \in \mathbb{R}^{K}$ is a vector of uplink powers.
    Finally,  $\rho_\text{max}(\cdot)$ denotes the spectral radius of a matrix, $\mathbf{D}_\mathbf{U} \triangleq \diag\left( \tfrac{|\mathbf{h}_1^\textsf{H} \mathbf{u}_1|^2}{\gamma_1}, \ldots, 
\tfrac{|\mathbf{h}_K^\textsf{H} \mathbf{u}_K|^2}{\gamma_K} \right) \in \mathbb{R}^{K \times K}$ is a diagonal matrix
and $\mathbf{F}_\mathbf{U} \in \mathbb{R}^{K \times K}$ is a matrix defined by
\begin{subnumcases}{\label{eq:F}\left[ \mathbf{F}_\mathbf{U} \right]_{ij} \triangleq } 
$0$, & for $i = j$ \\ 
|\mathbf{h}_i^\textsf{H} \mathbf{u}_j|^2, & for $i \neq j$
\end{subnumcases}
\begin{IEEEproof}
    The idea of the proof is rooted in the classic result of~\cite{bengtsson2018optimum}. We omit the details due to space limitations.
\end{IEEEproof}
\end{theorem}

Theorem~\ref{thm:ULDL_duality} states that whenever the pair $\left(\lambda, \boldsymbol{\beta}\right)$ is admissible, the downlink solution of problem~\eqref{prob:fixed_lambda} can be recovered by solving the uplink problem~\eqref{prob:UL_problem_fixed_lambda}.  More specifically, the optimal combiners for the uplink problem~\eqref{prob:UL_problem_fixed_lambda} are also optimal as beamforming directions for the downlink problem~\eqref{prob:fixed_lambda}.

Note that problem~\eqref{prob:UL_problem_fixed_lambda} differs from the standard uplink formulation, e.g.,~\cite{rashidUL1998}, in that it has the additional constraint~\eqref{eq:spectral_rad}. This constraint arises because 
the matrix $\lambda \mathbf{I} - \mathbf{Q}_{\boldsymbol{\beta}}$ does not need to be PSD in a global sense, and we only need $(\lambda, \boldsymbol{\beta}) \in \mathcal{A}$. 
If such matrix is already PSD then a simple proof~\cite{bengtsson2018optimum} reveals that $\rho_\text{max} \left( \mathbf{D}_\mathbf{U}^{-1} \mathbf{F}_\mathbf{U} \right) < 1$ is equivalent to $\mathbf{z} \geq 0$.
In which case, problem~\eqref{prob:UL_problem_fixed_lambda} reduces to the conventional uplink problem.
%

Observe that $(\lambda^*, \boldsymbol{\beta}^*)$ must be admissible since problem~\eqref{prob:fixed_lambda} is unbounded below whenever $(\lambda, \boldsymbol{\beta}) \not \in \mathcal{A}$, but the BCRB can never be negative.

\section{Proposed Algorithm}
The previous uplink interpretation is useful in devising efficient algorithms for solving the ISAC problem~\eqref{prob:generalCase}. Let us first focus on solving the uplink problem~\eqref{prob:UL_problem_fixed_lambda}.
If problem~\eqref{prob:UL_problem_fixed_lambda} is given by the classical formulation (i.e., with constraint~\eqref{eq:spectral_rad} replaced by $\mathbf{z} \geq 0$), the iterative procedure of~\cite{rashidDL1998} would recover an optimal uplink solution. Such iterations alternate between maximum SINR combining for fixed $\mathbf{z}$  
\begin{equation}\label{eq:UL_SINR_max}
    \mathbf{u}_k[n] \in \arg \min_{ \mathbf{u} \neq 0} \frac{\sum_{i \neq k}  z_i[n] \left| \mathbf{h}_i^\textsf{H}\mathbf{u}\right|^2 +  \mathbf{u}^\textsf{H} \left( \lambda \mathbf{I} - \mathbf{Q}_{\boldsymbol{\beta}} \right)\mathbf{u}}{\tfrac{1}{\gamma_k} \left|\mathbf{h}_k^\textsf{H} \mathbf{u} \right|^2},
\end{equation}
and a power control update for fixed $\mathbf{U}$
\begin{equation}~\label{eq:UL_power_update}
    z_k[n+1] = \frac{\sum_{i \neq k}  z_i[n] \left| \mathbf{h}_i^\textsf{H}\mathbf{u}_k[n]\right|^2 +  {\mathbf{u}^\textsf{H}_k[n]} \left( \lambda \mathbf{I} - \mathbf{Q}_{\boldsymbol{\beta}} \right)\mathbf{u}_k[n]}{\tfrac{1}{\gamma_k} \left|\mathbf{h}_k^\textsf{H} \mathbf{u}_k[n] \right|^2}.
\end{equation}  

It turns out that the same procedure, when initialized properly, can also recover the optimal solution of problem~\eqref{prob:UL_problem_fixed_lambda}.
\begin{theorem}\label{thm:alg_A_convergence}
        Let $\mathbf{z}^*_{\lambda, \boldsymbol{\beta}}$ and $\mathbf{U}^*_{\lambda, \boldsymbol{\beta}}$ denote an optimal uplink solution of problem~\eqref{prob:UL_problem_fixed_lambda}. If $(\lambda, \boldsymbol{\beta}) \in \mathcal{A}$, then starting from  $\mathbf{z}[0] \geq \mathbf{z}^*_{\lambda, \boldsymbol{\beta}}$, we have $\mathbf{z}[n] \rightarrow \mathbf{z}^*_{\lambda, \boldsymbol{\beta}}$ and $\mathbf{U}[n] \rightarrow \mathbf{U}^*_{\lambda, \boldsymbol{\beta}}$. 
        \begin{IEEEproof}
            Omitted due to space limitations.
        \end{IEEEproof}
    \end{theorem}
    
Note that the iterations 
converge to an uplink solution if $\left(\lambda, \boldsymbol{\beta} \right) \in \mathcal{A}$. If the pair is inadmissible, the decreasing sequence $\{\mathbf{z}[n]\}$ would fail to converge, which can be detected by checking if the sequence falls below zero.

After solving the uplink problem~\eqref{prob:UL_problem_fixed_lambda} for fixed $\left(\lambda, \boldsymbol{\beta}\right)$, a downlink solution can be obtained using~\eqref{eq:opt_V_fixed}. The problem of finding the optimal pair $\left( \lambda^*, \boldsymbol{\beta}^* \right)$ can be solved iteratively as an outer maximization of problem~\eqref{prob:partial_dual}, where each iteration involves solving an inner subproblem of the form~\eqref{prob:fixed_lambda} (see~\cite{yutransmitter2007}), 
along with a projection subgradient method onto $\mathcal{A}$ in the outer problem. Unfortunately, 
a projection operator on $\mathcal{A}$ cannot be readily obtained. Instead, we propose a simple scheme that works quite well. The idea is to update the parameters using the subgradient method (without projection onto $\mathcal{A}$) to obtain a tentative $(\tilde{\lambda}, \tilde{\boldsymbol{\beta}})$. If the tentative parameters are inadmissible, we reduce the step size by a factor and repeat the process until an admissible pair is found. When this simple scheme converges to the optimal pair, it would find the optimal beamformers $\mathbf{V}^*_{\lambda^*, \boldsymbol{\beta}^*}$ of the ISAC problem~\eqref{prob:generalCase}. 

\section{Simulations}
In this section, we examine the performance of the proposed beamforming solution for the angle of arrival problem in~\cite{LiuFCRB2022} against that of the SDR scheme. 
In this case, we set $N_\text{T} = N_\text{R} = 20$ antennas and $K = 2$ users with SINR thresholds of $10$ and $12$ dB. We model the communication channels as line-of-sight with angles $-30^\circ$ and $50^\circ$. We place the target at $0^\circ$ with unit path gain and choose a Gaussian prior with mean equal to the true values. 

\begin{figure}
    \centering    \includegraphics[width=0.45\textwidth]{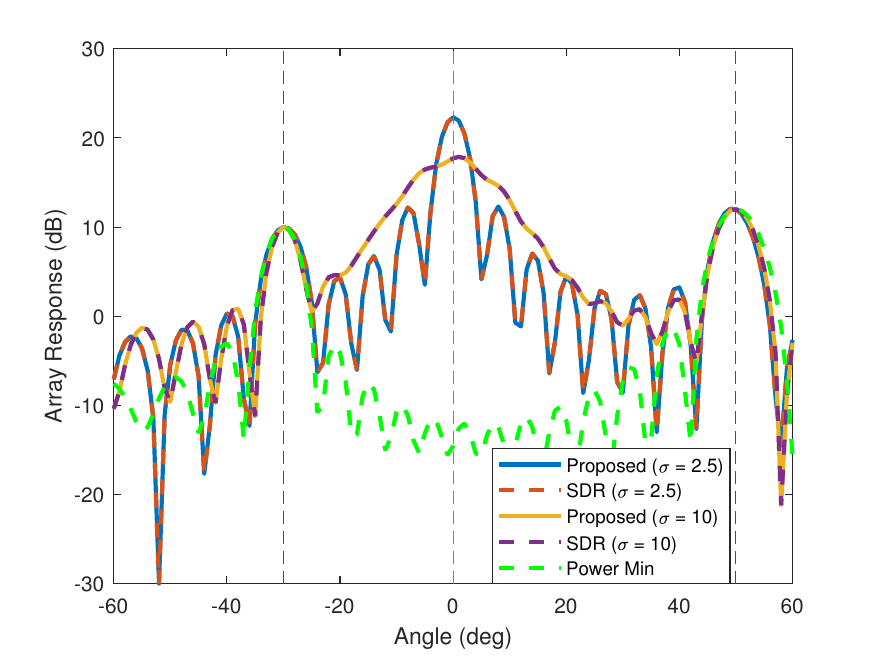}
     \caption{Beam pattern of the proposed solution vs SDR for optimizing the angle CRB. Here, we set $N_\text{T} = N_\text{R} = 20$ and $K = 2$.}
     \label{fig:comparison}
\end{figure}

In~\figurename~\ref{fig:comparison}, we plot the array response of the beamformers produced by both algorithms. We consider two distinct cases for the standard deviation of the prior distribution of the angle $\sigma = 2.5^\circ$ and $10^\circ$. In both cases, it is seen that the array response of the beamformer obtained by the proposed algorithm is identical to its SDR counterpart. However, the proposed method requires much a lower computational complexity. This highlights the benefits of the propose solution. We additionally observe that both algorithms form beams in the direction of communication users and the sensing target, but with a wider sensing beam when the variance of the prior is larger. Finally, we also plot the beam pattern for the beamformer that solves the classical communication problem~\eqref{prob:dl_comm} without considering sensing. We see that such beamformer can focus the beam towards the communication users but is clearly oblivious to the sensing target.

\section{Conclusion}
This paper develops a novel optimization framework for beamforming design in ISAC systems. Such framework is a consequence of the observation that a general optimization of a BCRB objective can be viewed as a maximization of beamforming power along with a set of auxiliary variables. We leverage this observation to develop an UL-DL duality for beamforming optimization in ISAC systems. This duality result allows the problem to be solved in the beamforming space at much lower computational complexity as compared to the existing SDR methods. 


\clearpage

\enlargethispage{-1.2cm} 
\IEEEtriggeratref{12}


\bibliographystyle{IEEEtran}
\bibliography{IEEEabrv,ref}
\end{document}